\documentclass[11pt,twoside]{article}
\usepackage{asp2010}
\usepackage{times}
\usepackage{epsfig}
\usepackage{natbib}
\usepackage{amssymb}
\usepackage{graphicx}
\usepackage{graphics}
\usepackage{pstricks}
\usepackage{rotating}
\usepackage{txfonts}
\usepackage{fancyhdr}
\usepackage{lscape}
\usepackage{calc}
\usepackage{float}                      
\usepackage{dcolumn}                    
\usepackage{multirow}
\usepackage{longtable}
\usepackage{subfigure}
\usepackage{appendix}
\usepackage{epstopdf}
\usepackage{array}

\def\be{\begin{equation}}\def\bea{\begin{eqnarray}}\def\beaa{\begin{eqnarray*}}
  \def\ee{\end{equation}}  \def\eea{\end{eqnarray}}  \def\eeaa{\end{eqnarray*}}

\def\fun#1#2{\lower3.6pt\vbox{\baselineskip0pt\lineskip.9pt
        \ialign{$\mathsurround=0pt#1\hfill##\hfil$\crcr#2\crcr\sim\crcr}}}

\newcommand{\an}{Astronomische Nachrichten}
\newcommand{\spie}{Society of Photo-Optical Instrumentation Engineers Conference Series}
\newcommand{\atel}{The Astronomer's Telegram}
\newcommand{\iaus}{IAU Symposium}

\newcommand{\basi}{Bulletin of the Astronomical Society of India}

\resetcounters

\begin{document}

\title{Spectroscopic and Photometric Development of T Pyxidis (2011) \\
    from 0.8 to 250 Days After Discovery}
\author{F. Surina$^1$, R. A. Hounsell$^2$, M. F. Bode$^1$, M. J. Darnley$^1$, D. J. Harman$^1$ and F. M. Walter$^3$}
\affil{$^1$Astrophysics Research Institute, Liverpool John Moores University, Twelve Quays House, Egerton Wharf, Birkenhead, CH41 1LD}
\affil{$^2$Space Telescope Science Institute, 3700 San Martin Drive, Baltimore, MD 21218 410-338-4974}
\affil{$^3$Department of Physics and Astronomy, Stony Brook University, Stony Brook, NY 11794-3800}

\begin{abstract}
We investigated the optical light curve of T Pyx during its 2011 outburst through compiling a database of SMEI and AAVSO observations. The SMEI light curve, providing unprecedented detail with high cadence data during $t$=1.5-49 days post-discovery, was divided into four phases based on the idealised nova optical light curve; the initial rise, the pre-maximum halt (or the \textquoteleft plateau\textquoteright), the final rise, and the early decline. Variation in the SMEI light curve reveals a strongly detected period of 1.44$\pm$0.04 days before the visual maximum. The spectra from the LT and SMARTS telescopes were investigated during $t$=0.8-80.7 and 155.1-249.9 days. The nova was observed very early in its rise and a distinct high velocity ejection phase was evident. A marked drop and then gradual increase in derived ejection velocities were present. Here we propose two different stages of mass loss, a short-lived phase occurring immediately after outburst followed by a more steadily evolving and higher mass loss phase. The overall spectral development follows that typical of a Classical Nova and comparison to the photometric behaviour reveals consistencies with the simple evolving pseudo-photosphere model of the nova outburst. The optical spectra are also compared to X-ray and radio light curves. Weak [Fe X] 6375$\AA$ emission was marginally detected before the rise in X-ray emission. The middle of the plateau in the X-ray light curve is coincident with the appearance of high ionization species detected in optical spectra and the peak of the high frequency radio flux.
\end{abstract}

\section{Introduction}

T Pyx has had six previous observed outbursts in 1890, 1902, 1920, 1944, 1966/1967 with an extensive nova shell associated with these. The outburst of T Pyx in 2011 was discovered by AAVSO observer M. Linnolt at a visual magnitude of 13.0 on 2011 Apr 14.29 UT (JD 2455665.7931, hereafter $t$=0 day) and published in \citet{sch11}. Other papers in these proceedings discuss this system - see e.g. those by Patterson et al. (optical), Ederoclite (optical and near IR), Chomiuk (radio), Mukai et al. (X-ray), and Orio (X-ray).


\section{Observations \& Results}
\subsection{Light curve from SMEI}
Photometric observations of T Pyx were obtained with the Solar Mass Ejection Imager (SMEI; \citealp{hou10,hou11} and Darnley et al. these proceedings). With its high-cadence all-sky observations, SMEI was able to investigate bright nova explosions ($m_{SMEI}$$<$8) whose outbursts occurred within the time period of operation (i.e. during 2003-2011) and produce extremely detailed light curves. The T Pyx SMEI light curve is compiled from 533 observation points and provides unprecedented detail with high cadence data and is compared to AAVSO light curves in Figure \ref{tpyx-smei-lc}.

We divide the SMEI light curve into 4 parts based on the idealised nova optical light curve given in \citet{war08} including the initial rise ($t$=1.5-3.3 days), the pre-maximum halt ($t$=3.3-13.3 days) hereafter the \textquoteleft plateau\textquoteright phase, the final rise ($t$=14.7-27.9 days), and the early decline ($t$=27.9-90 days). When comparing the SMEI light curve to the AAVSO light curves, we see that the nova seems to be bluer and similar to the magnitude in the $B$ filter at the plateau phase, then exhibits the same brightness in the $B$ and $V$  at the visual maximum where it is expected to behave like an A0 star, and tends to be redder, approaching the magnitude in the $R$ filter, after the visual maximum as shown in Figure \ref{tpyx-smei-lc}.

\subsubsection{Investigation of Early Time Periodicity}
The apparently aperiodic stochastic or so-called flickering SMEI light curve was search-ed for any periodic modulations. The analysis was made throughout all the available SMEI data but separated into 4 cases which are (a) from the first observation to the last observation, (b) from the first observation to the visual maximum, (c) from the visual maximum to the last observation, and (d) from the first observation to the end of the plateau phase. They were analysed with the $PERIOD04$ PC code from \citet{len04,len05}. Case (d) yields a strongly detected period $P$=1.44$\pm$0.04 days. This period is close to the weak signal of 1.24 days found by \citet{pat98} who suggested it might originate from precession in the accretion disk. With a mass ratio of $q$=0.2 for T Pyx \citep{uth10}, the work of \citet{hir90} on typical CVs (where the accretion disk can develop a non-axisymmetric structure and finally settles into a periodically oscillating state) suggests that the ratio of $P_{orb}/P_{precession}$ should be 7.7$\%$. Our observed $P_{orb}/P_{precession}$ is 5.3$\%$ if the observed oscillations are indeed due to disk precession. Although this is similar to the explanation from \citet{hir90}, it is doubtful whether a significant disk will be present and observable at this time and therefore the cause of these oscillations remains uncertain.




\subsection{Spectra from LT and SMARTS}
\begin{figure} [tp!]
\centering
\begin{minipage}{.5\textwidth}
  \centering
  \includegraphics[width=16pc]{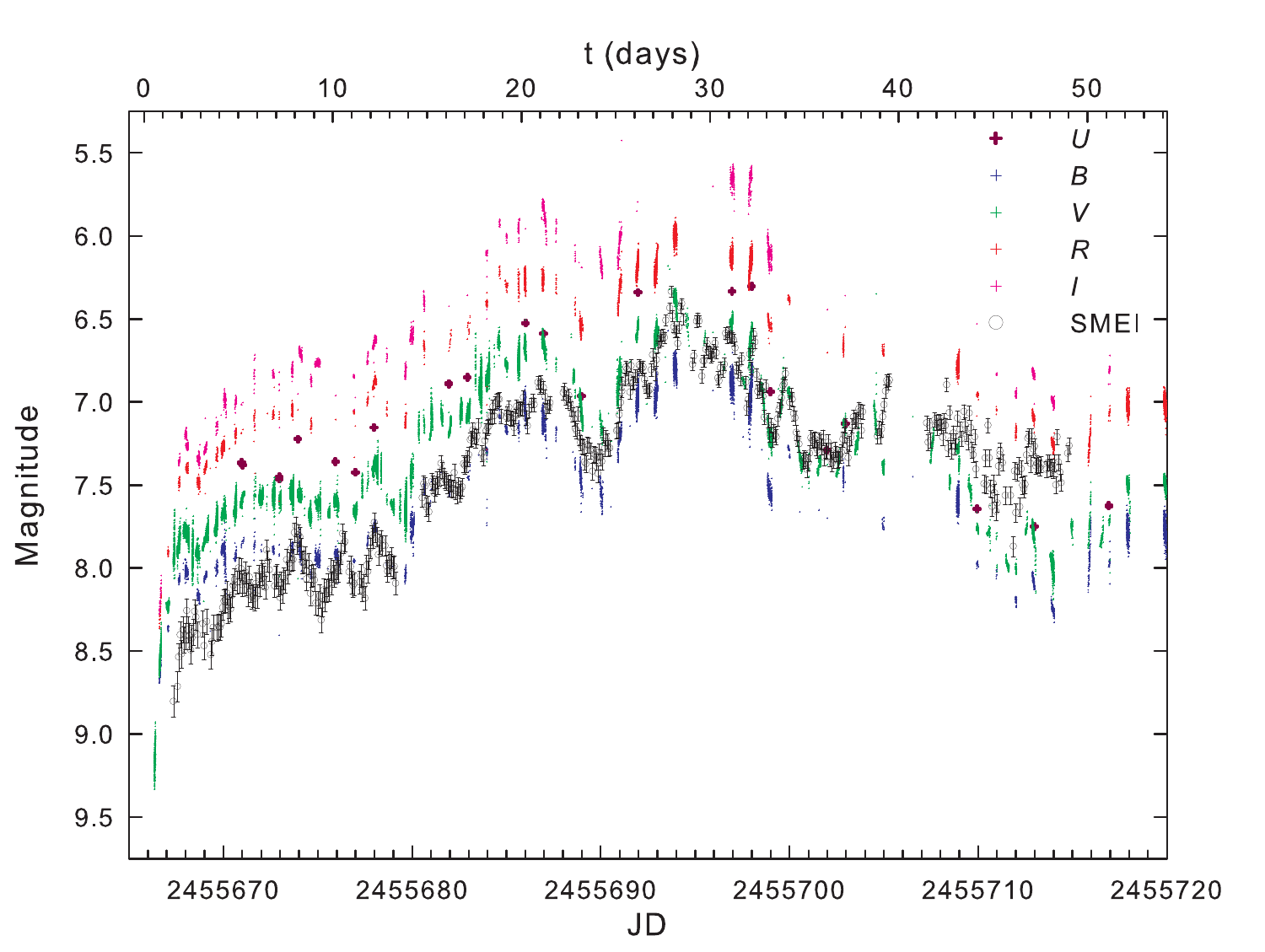}
  \caption{The SMEI light curve of T Pyx during its 2011 outburst (open black circles with error bars) compared to $UBVRI$ light curves observed by AAVSO (plus signs). Note that colour versions are available via astro-ph arXiv1303.6592.}
  \label{tpyx-smei-lc}
\end{minipage}%
\begin{minipage}{.5\textwidth}
  \centering
  \includegraphics[width=12pc]{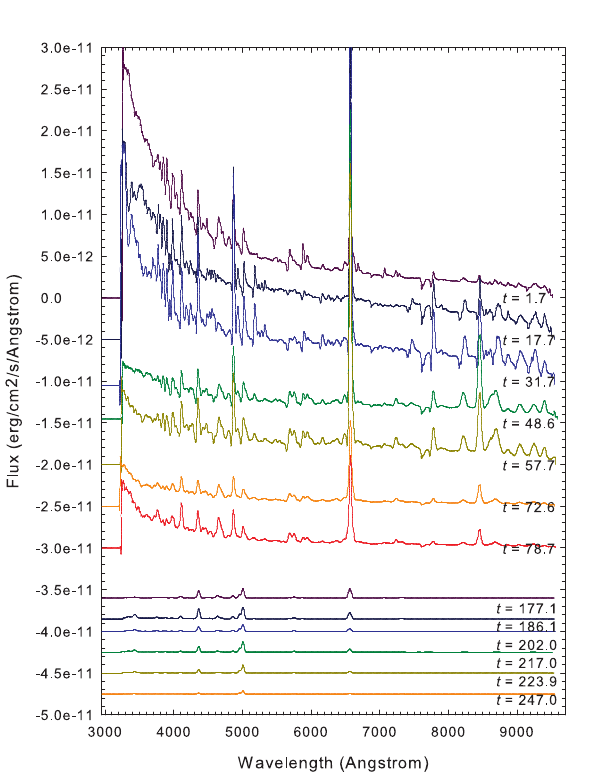}
  \caption{Low-resolution spectra of T Pyx taken by the 1.5m SMARTS telescope from $t$=1.7-247 days.}
  \label{all-low-res}
\end{minipage}
\end{figure}

Spectroscopic observations were carried out by using the 1.5m telescope of the SMARTS II Consortium located at CTIO Chile and equipped with a long-slit R-C spectrograph. We obtained 99 low-to-moderate-resolution (300 $<$ R $<$ 3400) optical spectra of T Pyx from $t$=0.8-80 days and from  from $t$=155.1-249.9 days from the SMARTS atlas \citealp[see][]{wal12}. The 13 low-resolution spectra from SMARTS are presented in Figure \ref{all-low-res}. The 2m robotic Liverpool Telescope (LT; \citealp[see][]{ste04}) sited at the Observatorio del Roque de Los Muchachos on the Canary Island of La Palma, Spain also secured spectra from $t$=8.6-11.6 days using the Fibre-fed RObotic Dual-beam Optical Spectrograph (FRODOSpec; \citealp[see][]{bar12}). 

The optical light curve was compared both to the flux and to the ejection velocities measured from H$\alpha$, H$\beta$, and H$\gamma$ lines as shown in Figure \ref{lc-vej-flux-H-FeII}. The variation in flux of the Balmer lines is similar to that found in the light curves. We note the H$\alpha$ \textquoteleft flare\textquoteright at $t$=33.7 days which may be responsible for the sharp and high-amplitude peak in the variation of the light curve at this time.

\subsubsection{Spectral evolution}

Here we discuss the spectral evolution based on the idealised nova optical light curve given in \citet{war08} together with the recognition of a common pattern of line development for CNe described by \citet{mcl42,mcl44}. Figure \ref{all-low-res} shows all low-resolution spectra taken from $t$=1.7-247 days which cover the initial rise phase through the transition phase of the light curve. The flux and the ejection velocities ($V_{ej}$, derived from P Cygni profiles) of the Balmer lines, the Fe II recombination lines at 5169$\AA$, 5018$\AA$, 4233$\AA$, 4178$\AA$, 4173$\AA$, the O I 7775$\AA$, and the Ca II K line at 3934$\AA$ are shown in Figure \ref{lc-vej-flux-H-FeII}.

\begin{enumerate}
\item {\itshape The initial rise ($t$=0.8-3.3 days):\/} The measured flux shows that the emission comes almost entirely from the continuum. There is also a marked drop in $V_{ej}$ from $\sim$4000 km s$^{-1}$ at $t$=0.8 days to $\sim$2000 km s$^{-1}$ at $t$=2.7 days. This drop is similar to that seen in DQ Her \citep{mcl37} and V603 Aql \citep{wys40}. If the initial ejection is a Hubble flow, then one will a see high apparent $V_{ej}$ initially, declining as one sees into deeper layers as  the mass loss rate decreases. We also note as an aside comparison to the high velocity features (HVFs) found in all Type Ia SNe and believed to be the result of the interaction of initial highest velocity ejecta with a circumstellar envelope \citep{ben05}. The subsequent change in behaviour of the derived $V_{ej}$ during the plateau phase suggests 2 different stages of mass loss: a short-lived phase first occurring immediately after outburst and then followed by a more steadily evolving and higher mass loss phase.
\begin{figure}
\begin{center}
\leavevmode
\epsfxsize = 14.0cm
\epsfysize = 14.0cm
\includegraphics[width=20pc]{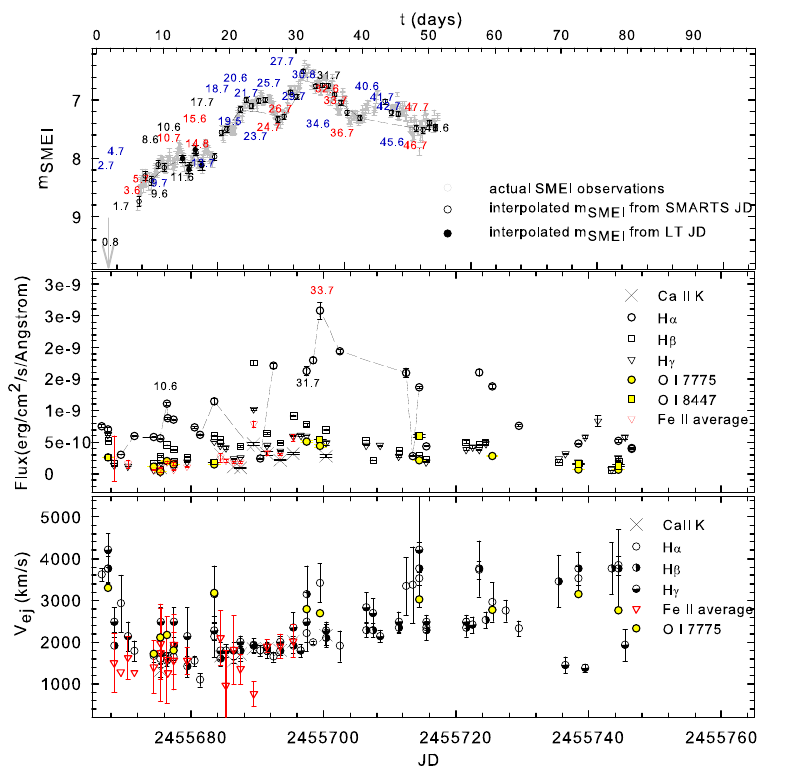}
\end{center}
\caption[SMEI light curve of T Pyx compared to the measured line fluxes and the ejection velocity evolution (see text for details).]
{SMEI light curve of T Pyx compared to the measured line fluxes and the ejection velocity evolution.}
\label{lc-vej-flux-H-FeII}
\end{figure}

\item {\itshape Plateau ($t$=3.6-13.7 days):\/} The increasing intensity of Fe II and O I emission lines and the change in spectral type from B to A suggest the \textquoteleft $iron$ $curtain$\textquoteright stage \citep{schw01} begins around $t$=8.6 days. Light curve variation seems to be consistent with the line strengths of H$\alpha$, H$\beta$, H$\gamma$, and Fe II. Meanwhile the derived $V_{ej}$ decreases and starts to stabilise at$\sim$1500 km s$^{-1}$ during this halt.

\item {\itshape Final rise ($t$=14.7-27.9 days):\/} The O I flash begins at $t$$\sim$17 days. The typical principle spectrum is apparent while the iron curtain still persists. The major dip in the SMEI light curve at $t$=23.7 days appears to be caused by the same mechanism that makes the flux in H$\alpha$ drop significantly. The light curve then rises more steeply toward the maximum at $t$=27.9 days. Here $V_{ej}$ is gradually increasing. This implies that the relative size of the pseudo-photosphere to the ejecta is shrinking and revealing higher velocity material.

\item {\itshape Early decline ($t$=27.9-90 days):\/} The Principle spectrum persists until $t$$\le$48 days. Balmer lines begins to show double-peaked structure at $t$=42.7 days while the forbidden line [O III] 5007$\AA$ begins to emerge at $t$=45.6 days. The typical Orion spectrum is obvious at $t$$\sim$79 days due to the strong N III + N II 4640$\AA$ emission. Here weak He I 5876$\AA$ and [Fe X] 6375$\AA$ were marginally detected at $t$$\sim$80 days which is the last spectrum observed before the seasonal gap. Meanwhile the pseudo-photosphere is continuously shrinking and the effective temperature is increasing as a result. Ultimately the envelope becomes optically thin.

\item {\itshape Transition phase ($t$=90-250 days):\/} The first spectrum observed after the seasonal gap at $t$=155.1 days shows that the nova has evolved into the nebular stage.
\end{enumerate}

Throughout, we found that T Pyx generally followed the simple pseudo-photosphere model where the properties of all nova spectra are expected to be similar at the same magnitude below peak \citep{bat89}. Figure \ref{radio-x-lc} (top panel) shows where different emission lines were expected to be evident in this model.

\section{Comparison to the X-ray and Radio Light Curves}
\begin{figure}
\begin{center}
\leavevmode
\epsfxsize = 14.0cm
\epsfysize = 14.0cm
\includegraphics[width=30pc]{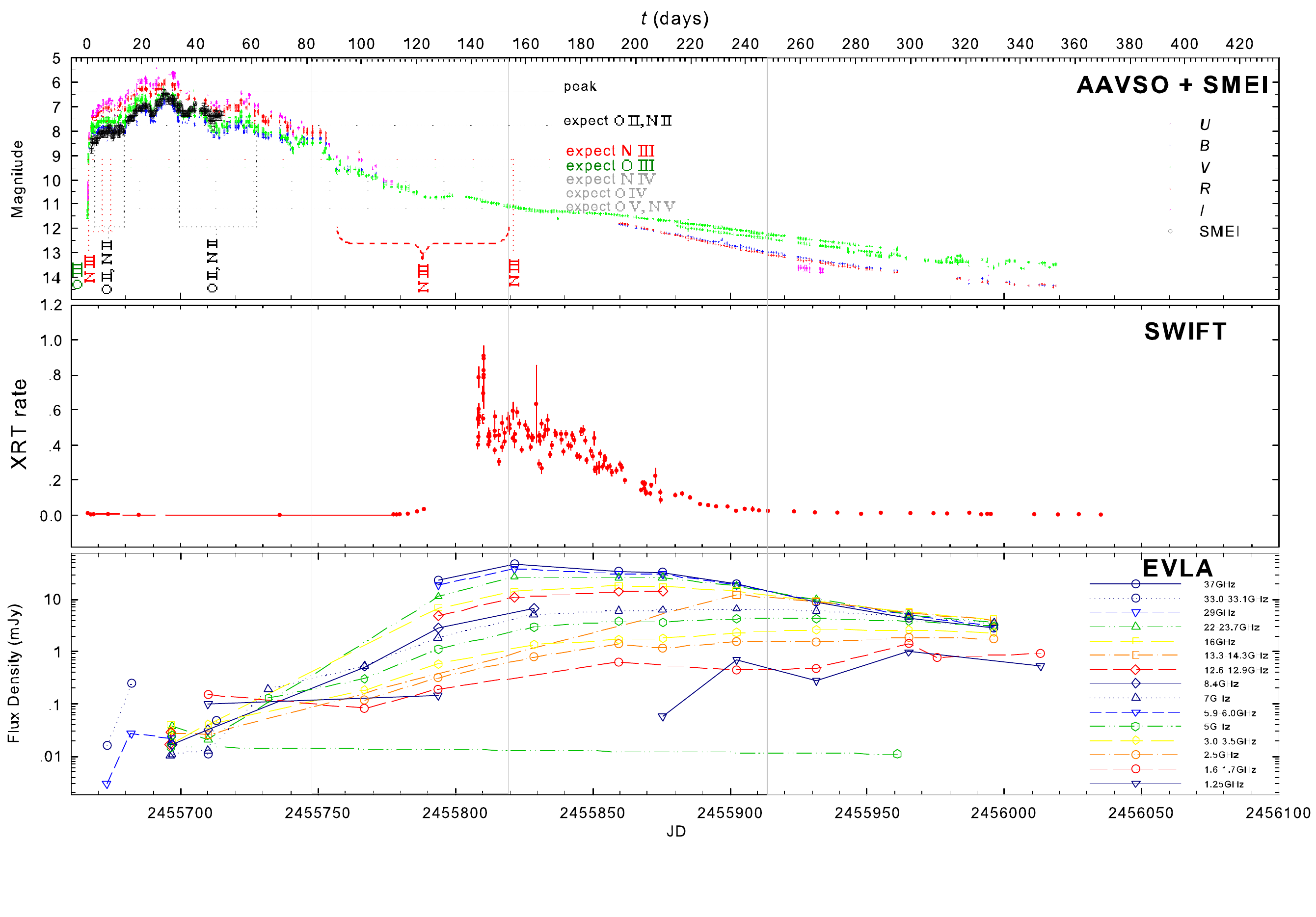}
\end{center}
\caption[SMEI light curve of T Pyx compared to measured line fluxes and ejection velocity evolution (see text for details).]
{SMEI light curve of T Pyx compared to measured line fluxes and ejection velocity evolution (see text for details).}
\label{radio-x-lc}
\end{figure}

We compared the optical light curves to the X-ray light curve from SWIFT and radio light curves from the EVLA \citep{nel12} as shown in Figure \ref{radio-x-lc}. The rise of high frequency radio emission (37 GHz) began at $t$=7-15 days during the plateau phase in optical light curve. The radio emission then rose steeply during the early decline where [O III] 5007$\AA$ was first present at $t$$\sim$45 days. The X-ray emission was detected at $t$=111 days and peaked at $t$=144 days which is consistent with the appearance of [Fe X] 6375$\AA$ found marginally earlier (at $t$=80 days) and high ionization species found later (at $t$=155.1 days). We note that the Chandra grating spectra of this phase showed that emission lines were very strong in the X-ray spectrum \citep{ori12}. The X-ray emission may appears to be a mix of super soft source and shocked circumstellar gas. The middle of the plateau in the X-ray light curve ($t$$\sim$155 days) is consistent with the time when the high frequency (37 GHz) radio flux peaked and the presence of the coronal lines [Fe X] 6375$\AA$ and [Fe VII] 6087$\AA$ in the optical spectra ($t$=161.1 days). X-ray emission was finally undetectable at $t$$\sim$222 days but the low frequency radio (1.25 GHz) reached a peak later at $t$$\sim$250 days.

\section{Conclusions}
The SMEI light curve provided unprecedented detail with high cadence data of the outburst of T Pyx in 2011 from $t$=1.5-49 days. A period of 1.44$\pm$0.04 days was detected before visual maximum. The spectra from LT and SMARTS were obtained during $t$=0.8-80 days and 155-250 days. The nova was observed spectroscopically very early in its rise and a distinct high velocity ejection phase was evident at this time.
The overall spectral development followed that typical of a Classical Nova and comparison to the photometric behaviour revealed consistencies with the simple evolving pseudo-photosphere model of the nova outburst. Weak [Fe X] 6375$\AA$ was marginally detected before the rise in X-rays. The middle of the plateau in the X-ray light curve was consistent with the appearance of high ionization species and the peak of high frequency radio flux. A full account will appear in Surina et al. (in preparation).

\acknowledgements
We Thank Kim Page for supplying the XRT light curve which was created from observations obtained
by the Swift nova-CV group (http://www.swift.-ac.uk/nova-cv/) and Laura Chomiuk for supplying the EVLA radio data.

\clearpage

\end{document}